\documentclass[
 reprint,
superscriptaddress,
 amsmath,amssymb,
 aps,
pra,
longbibliography
]{revtex4-1}

\usepackage{siunitx}
\usepackage{graphicx}
\usepackage{epsfig}
\usepackage{float}
\usepackage{lipsum}
\usepackage[margin=2cm,includefoot]{geometry}
\usepackage[colorlinks, linkcolor=black]{hyperref}
\usepackage[english]{babel}
\usepackage{fancyhdr}
\usepackage{subfiles}
\usepackage{amsthm}
\usepackage{amsfonts}
\usepackage{braket}
\usepackage{endnotes}
\usepackage{physics}
\usepackage{blindtext}
\usepackage{cleveref}
\usepackage{color}

\begin{document}

\title{Superresolution imaging of two incoherent sources via two-photon interference sampling measurements in the transverse momenta}

\author{Salvatore Muratore}
\affiliation{School of Mathematics and Physics, University of Portsmouth, Portsmouth PO1 3QL, UK}

\author{Danilo Triggiani}
\affiliation{School of Mathematics and Physics, University of Portsmouth, Portsmouth PO1 3QL, UK}
\affiliation{Dipartimento Interateneo di Fisica, Politecnico \& Universit\`a di Bari, Bari 70126, Italy}

\author{Vincenzo Tamma}
\email{vincenzo.tamma@port.ac.uk}
\affiliation{School of Mathematics and Physics, University of Portsmouth, Portsmouth PO1 3QL, UK}
\affiliation{Institute of Cosmology and Gravitation, University of Portsmouth, Portsmouth PO1 3FX, UK}

\date{\today}

\begin{abstract}
The Rayleigh's criterion infamously imposes a minimum separation between two incoherent sources for them to be distinguishable via classical methods. In this work, we demonstrate the emergence of two-photon beats from the interference of a single reference photon and a photon coming from one of two transversally displaced incoherent sources. We also show that, apart from a factor of two, the ultimate quantum precision in the estimation of any value of the distance between two thermal sources is achievable independently of the wavepacket spatial structure, by performing a relatively low number of sampling measurements of the transverse momenta of the interfering photons, without the need of any additional optics. The feasibility of this technique makes it an optimal candidate to important applications in microscopy, astronomy and remote sensing.
\end{abstract}

\maketitle

\paragraph*{Introduction}
The estimation of the position of two close incoherent sources is one of the most notorious problems in optics. Indeed the Rayleigh criterion defines the minimum separation between the two sources that can be measured by means of classical measurements such as direct imaging, due to diffraction~\cite{BornWolf1999}. On the other hand, the possibility to perform sub-diffraction imaging is essential in those scenarios where two or more incoherent sources are close within the diffraction limit, e.g. in biomedical imaging and nanoscopy~\cite{Sahl2017, Khater2020}, and astronomy~\cite{Sandler1994}.  
The research of new methods of measurements in the field of imaging and sensing of thermal sources in the sub-diffraction regime has thus recently stimulated a thriving and successful quest to break this limit, for long though to be fundamental. 

Indeed, different techniques have been developed in the past years, achieving a superresolution regime in the estimation of the separation between the sources.
Many of these techniques require a certain degree of control on the emission of the sources, e.g. as in stimulated emission depletion microscopy~\cite{Klar1999, Vicidomini2018} or blinking microscopy~\cite{Burnette2011}, which increases the complexity of the measurement, and certainly renders these techniques not applicable to astronomical observations.
One of the precursors of superresolution techniques applicable to astronomy is the spatial-mode demultiplexing method developed by Tsang et. al.~\cite{Tsang2016}, consisting in a clever decomposition of the wavepackets of the incoming thermal photons in orthogonal spatial modes.
However, despite the remarkable results achieved with this approach, the spatial decomposition of the signal depends on the particular spatial structure of the incoming wavepackets of the photons, and ultimately increases the complexity of the experimental apparatus, including a limitation in the sensitivity given by the mode crosstalk~\cite{Tsang2016, Paur2016, Linowski2023, Santamaria2023, Santamaria2024}.

\begin{figure}
\centering
\includegraphics[width=.85\columnwidth]{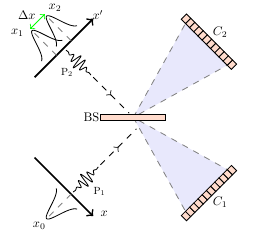}
\caption{Two-photon interferometer. A single photon $P_{2}$ coming from two thermal sources with relative transverse displacement $\Delta x$ interferes on a balanced beam splitter with a photon $P_{1}$ produced by a reference source. Two cameras $C_{1}$ and $C_{2}$ in the far field resolve the transverse momenta of the two photons, while simultaneously recording whether the photons hit the same camera or opposite cameras. By collecting a relatively small number of sampling events at the output it is possible to optimally retrieve information on the separation also when $\Delta x\simeq 0$.}
\label{fig:Setup}
\end{figure}
Some helpful insights in the development of new sensing schemes can be found in the fairly recent progresses in the field of two photon quantum interference applied to quantum sensing, referring in particular to the exploitation of the Hong-Ou-Mandel (HOM) effect~\cite{Hong1987, Shih1988, Lyons2018, Harnchaiwat2020, Sgobba2023}.
The HOM effect is a quantum effect for which, when two identical photons impinge on the faces of a balanced beam splitter, the two photons will always be observed bunching together at the same outputs ports. If the photons are not identical, e.g. in their polarization or emission times, etc., we will instead estimate coincidences with some non-vanishing probability. However such schemes present experimental challenges, such as the need of adaptation techniques, which limit the quantum enhancement in sensitivity in real world scenarios \cite{Triggiani2023, Triggiani2024}. More recently, techniques based on multiphoton interference with inner variables sampling, first used as scalable tools for demonstrating quantum computational complexity  ~\cite{ tamma2015multiboson, wang2018experimental, laibacher2018symmetries, tamma2014multiboson}, have allowed to overcome these challenges  in the quantum-enhanced estimation of photonic parameters, such as time delays, polarisations and displacements \cite{Triggiani2023, Triggiani2024, Maggio2024, legero2004quantum, legero2003time}.
Two-photon spatial interference has been also proposed towards the goal of achieving superresolution of thermal sources, although relying on the challenging task to store two thermal photons before making them interfere ~\cite{ parniak2018beating}. This would require, for example, the experimentally demanding need of photon number quantum nondemolition measuring devices with spatially multimode quantum memories ~\cite{ parniak2018beating}.

In this work we demonstrate the emergence of quantum beats from the interference of a photon coming form one of two transversally displaced thermal sources and a reference photon, which can open new routes for quantum optics experiments at an astronomical scale ~\cite{Deng2019}. Furthermore we propose a quantum superresolution technique based on transverse-momentum measurements of the two interfering photons, which overcome the experimental challenges previously mentioned in current methods.
In particular, after impinging on the two faces of a beam splitter, the two photons are then collected by two cameras in the far field where it is possible to resolve their transverse momenta.
Our technique does not require further optics to spatially demultiplex the thermal signal and is independent of the shape of the spatial wavepacket of the photons.
Moreover we will show that our scheme achieves constant quantum superresolution precision in the estimation of the separation between the two thermal sources independently of the value of the separation, which makes it useful when tracking a relative distance that can vary both inside and outside the sub-diffraction region.
Finally, we also show that we can further simplify the measurement scheme replacing the transverse-momentum resolving cameras with simple bucket detectors and still achieve the same superresolving precision for the estimation of small separations between the sources.\\

\paragraph*{Two-photon quantum interference with two transversally displaced thermal sources.}

In Fig.~\ref{fig:Setup}, we consider  two weak thermal sources separated transversally by an unknown distance $\Delta x$ and emitting longitudinally quasi-monochromatic photons with wavenumber $K_0$.
For simplicity we will limit the discussion to one transverse dimension.
The assumed low intensity $\varepsilon\ll 1$ of the sources allows us to describe their quantum state as 
\begin{equation}
\hat{\rho}=(1-\varepsilon)\hat{\rho}_{0}+\varepsilon\hat{\rho}_{1}+o(\varepsilon^{2})
\end{equation}
where $\hat{\rho}_{0}=\ket{\mathrm{vac}}\bra{\mathrm{vac}}$ denotes the zero photons state, $o(\varepsilon^{2})$ is the negligible contribution from higher number of photons, while\begin{gather}
\begin{gathered}
\hat{\rho}_{1}=\frac{1}{2}\left(\ket{\psi_{1}}\bra{\psi_{1}}+\ket{\psi_{2}}\bra{\psi_{2}}\right),\\
\ket{\psi_{i}}=\int^{\infty}_{-\infty}\dd x\ \psi(x-x_{i})\hat{a}_{1}^{\dagger}(x)\ket{\mathrm{vac}}, \quad i=1,2
\end{gathered}.
\label{eq:SourcePhoton}
\end{gather}
Here, $\hat{\rho}_1$ is the single-photon contribution, mixture of two single photon states with wavepacket $\psi(x-x_i)$, each centred at the imaged position $x_i$ of the $i$th source, with $i=1,2$, while $\hat{a}_1^\dag(x)$ is the bosonic creation operator associated with a photon at position $x$.
We notice that, in general, $\braket{\psi_1}{\psi_2}\neq 0$, for $x_{1}\neq x_{2}$.

We let the photon originated by the thermal sources impinge on one of the faces of a balanced beam splitter, while at the other input of the beam splitter we inject a single reference photon 
\begin{equation}
\ket{\psi_{0}}=\int_{-\infty}^{\infty}\dd x\ \psi(x-x_{0})\hat{a}^{\dagger}_{0}(x)\ket{0}, 
\label{eq:RefPhoton}
\end{equation}
with wavepacket $\psi(x-x_0)$ centred in $x_0$, where $\hat{a}_0^\dag(x)$ is the bosonic creator operator associated with a reference photon in position $x$.

Finally, after propagating for a distance $d$, the two photons are detected by two single-photon cameras positioned at the two output ports of the beam splitter in the far-field regime.
This allows us to detect the transverse momenta $k=y \frac{K_0}{d}$ and $k'=y' \frac{K_0}{d}$ of the photons, as these are proportional to the positions $y$ and $y'$ of the clicking pixels of the cameras, and the wavenumber of the photons $K_{0}$.
The outcome $(\Delta K,X)$ of a single measurement is given by the difference $\Delta K=k-k'$ in the transverse momenta of the detected photons, and whether they hit the same camera ($X=B$, bunching event) or opposite cameras ($X=A$, antibunching event). 
\begin{figure}
\includegraphics[width=.95\columnwidth]{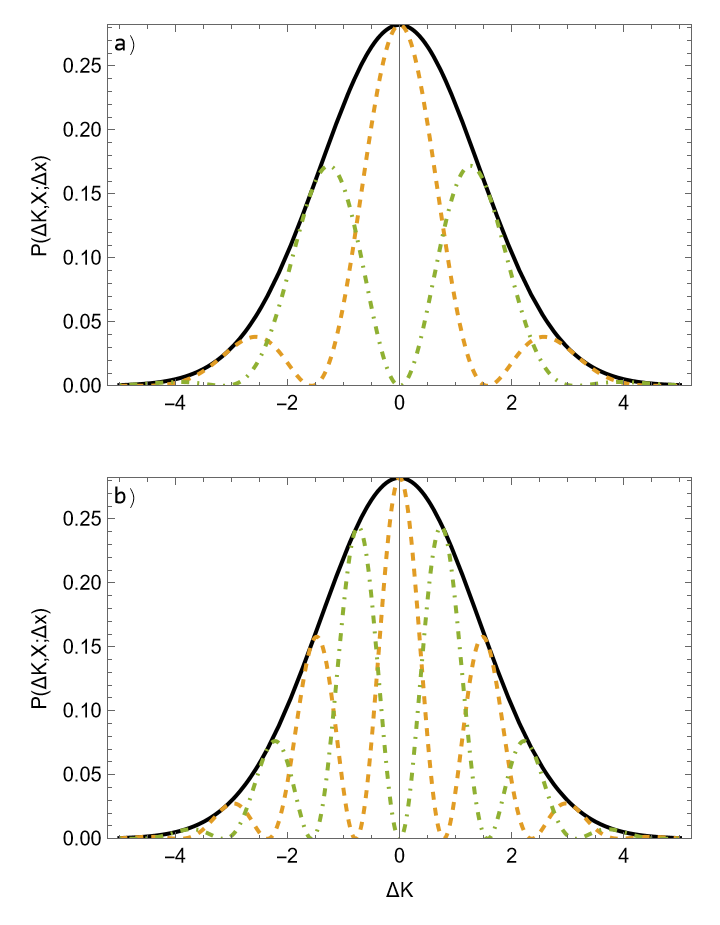}
\caption{Plots of the probability $P(\Delta K,X)$ in Eq.~\eqref{Eq:deltafixed} as function of the difference of transverse momenta $\Delta K$, considering as an example a Gaussian spatial wavepacket $\psi(x)$ with unitary variance and a) $\Delta x=4/\sigma_{k}$, b) $\Delta x=8/\sigma_{k}$. The probability of a bunching event ($X=B$), in dashed orange line, and the probability of a coincidence event ($X=A$), in dashed-dotted green line, manifest a quantum beat periodicity inversely proportional to the value of $\Delta x$.
}
\label{fig:Probs}
\end{figure}
In Appendix~\ref{app:Prob} we evaluate the probability 
\begin{multline}
P(\Delta K, X)=\frac{1}{2}C(\Delta K)\\
\times\bigg(1+\alpha(X) \cos(\Delta K\frac{\Delta x}{2})\cos(\Delta K\left(x_0-x_s\right))\bigg),
\label{eq:Prob}
\end{multline}
with $\alpha(B)=1,\, \alpha(A)=-1$, to observe any given pair of photons in a sampling outcome $(\Delta K,X)$~\cite{Suppl}. Such a probability depends on the separation $\Delta x=x_{1}-x_{2}$ and the centroid $x_s=(x_{1}+x_{2})/2$ of the two sources. Interestingly, it manifests two-photon interference beats in the variable $\Delta K$, with an envelope $C(\Delta K)$ depending on the Fourier transform of the position wavepacket $\psi(x)$, e.g. $C(\Delta K)=\exp(-\Delta K^{2}/4\sigma_{k}^{2})/\sqrt{4\pi\sigma_{k}^{2}}$ in the case of a Gaussian spatial distribution $\abs{\psi(x)}^2$ with variance $\sigma_x^2=1/4\sigma_k^2$.
For a separation $\Delta x=0$ between the two thermal sources, two-photon interference beatings in the probability in Eq.~\eqref{eq:Prob} as a function of $\Delta K$ can be observed with a periodicity $T=2\pi/(x_{0}-x_{s})$ inversely proportional to the transverse separation between the centroid position $x_{s}$ and the position $x_{0}$ of the reference photon. On the other hand, for thermal sources with a finite separation $\Delta x\neq 0$, one can align the interferometer to the centroid position which can be known through standard interferometric techniques ~\cite{Tsang2016}. In such a case, quantum beats in the probability
\begin{equation}
P(\Delta K, X)=\frac{1}{2}C(\Delta K)\bigg(1+\alpha(X) \cos(\Delta K\frac{\Delta x}{2})\bigg),
\label{Eq:deltafixed}
\end{equation} 
obtained from Eq.~\eqref{eq:Prob} for $x_{0}=x_{s}$, can be observed with a periodicity $T=4\pi/\Delta x$ inversely proportional to the value of $\Delta x$, as illustrated in Fig.~\ref{fig:Probs}. An important question spontaneously arises: can such an interferometer be used as a quantum sensor for the estimation of $\Delta x$?\\

\paragraph*{Quantum estimation technique}

We will now introduce a quantum metrological technique based on the two-photon quantum beat phenomena described before to estimate efficiently the separation $\Delta x$ by sampling from the probability distribution in Eq.~\eqref{Eq:deltafixed}. The estimation of $\Delta x$, is performed by measuring the sampling outcomes $\{(\Delta K_{i}, X_{i})\}_{i=1,\dots,N}$ of $N$ two-photon interference measurements  with the interferometer in Fig.~\ref{fig:Setup}  and evaluating the maximum-likelihood estimator $\widetilde{\Delta x}$ from such outcomes.
In particular we now show that our technique achieves superresolution, i.e. it is capable to accurately estimate the distance $\Delta x$ between two thermal sources beyond the Rayleigh criterion.
To do so, we evaluate the Cram\'er-Rao bound~\cite{Cramer1999}
\begin{equation}
\mathrm{Var}[\widetilde{\Delta x}]\geq \frac{1}{NF(\Delta x)}
\label{eq:CRB}
\end{equation} 
associated with our scheme, i.e. the uncertainty $\mathrm{Var}[\widetilde{\Delta x}]$ achievable by our metrological technique through maximum-likelihood estimation, where $N$ is the number of detected pairs and $F(\Delta x)$ is the Fisher information associated with momentum resolved measurements in the far field. 
\begin{figure}
\includegraphics[width=.95\columnwidth]{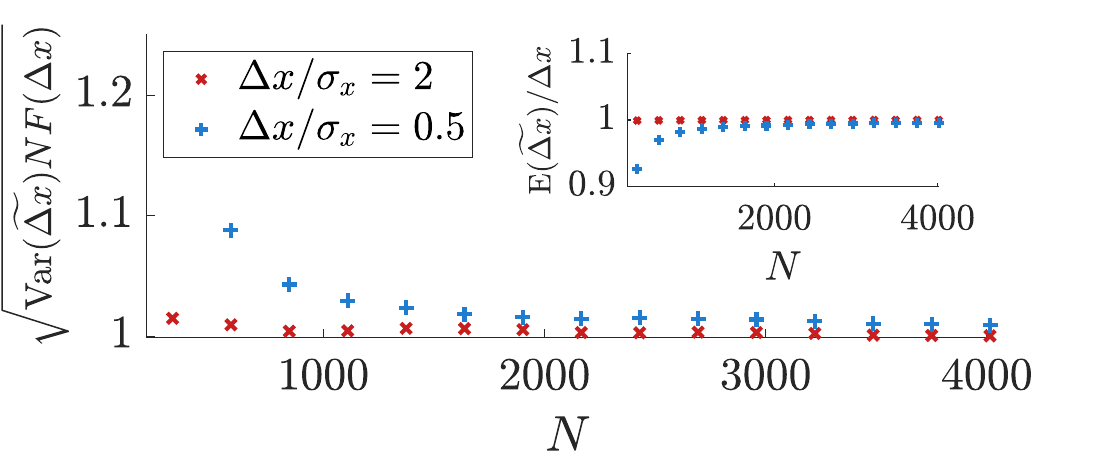}
\caption{Plots of the uncertainty associated with the maximum-likelihood estimator normalised to the Cram\'er-Rao bound in Eqs.~\eqref{eq:CRB} and ~\eqref{eq:Fisher}, estimated numerically for different values of number $N$ of detected pairs of photons and for two different values $\Delta x/\sigma_x=0.5,2$ inside and outside the subdiffraction regime, respectively. A saturation $\geq 99\%$ of the Cram\'er-Rao bound is reached already for $N=2000$. In the inset, we plot the estimated expectation value of the maximum-likelihood estimator normalized by its real value. The unbiasedness of the estimator emerges again already for $N=2000$.}
\label{fig:Numeric}
\end{figure}
Notoriously, the Fisher information associated with the estimation of $\Delta x$ through direct measurement tends to zero for $\Delta x\rightarrow 0$, implying the impossibility to estimate $\Delta x$ due to the Rayleigh criterion~\cite{Tsang2016}. On the
other hand we will show that our superresolution technique
is characterised instead by a constant Fisher information
even for $\Delta x\rightarrow 0$.

In Appendix~\ref{app:FI} evaluate the Fisher information found in Eq.~\eqref{eq:CRB} for the probability distribution in Eq.~\eqref{Eq:deltafixed} as
\begin{equation}
F(\Delta x)=\frac{\sigma^{2}_{k}}{2},
\label{eq:Fisher}
\end{equation}
where $\sigma_k^2$ is the variance of the transverse-momentum distribution of the photons \cite{Suppl}.
Noticeably, $F(\Delta x)$ is independent of $\Delta x$, meaning that our two-photon correlation-based scheme achieves the same precision and differs from the Quantum Fisher Information only by a constant factor $\frac{1}{2}$ indipendently of the spatial structure of the photonic wavepackets ~\cite{Tsang2016}.
The observable to be measured in our scheme, i.e. the second order correlations in the transverse momenta of the two interfering photons, is independent of the shape of the wavepacket $\psi(x)$ of the photons.
This feature renders our technique more feasible than recently proposed superresolution imaging techniques, which require to demultiplex the signal of the photon arriving from the thermal sources into orthogonal optical modes that depend on the wavepacket $\psi(x)$ of the photons requiring the use of a specific waveguide for each different kind of wavepacket, and the need of overcoming mode crosstalk ~\cite{Tsang2016, Paur2016, Santamaria2023, Santamaria2024}.\\
Moreover, for photons with significantly overlapping wavepackets, $\sigma_{k}\Delta x\ll 1$,  we show in Appendix~\ref{app:Nonres} that one can still obtain the same value of the Fisher Information in Eq.~\eqref{eq:Fisher} without the need to resolve transverse momenta, but using instead simple bucket detectors \cite{Suppl}.\\
Remarkably, in Fig.~\ref{fig:Numeric} we show the results of numerical simulations of the maximum-likelihood estimation, which indicate its unbiasedness and a 99\% saturation of the Cram\'er-Rao bound in Eq.~\eqref{eq:CRB} already for $N=2000$ pairs of photons. \\

\paragraph*{Conclusions}

In this work we introduce an interferometric scheme for the observation of two-photon quantum beat interference, based on transverse momentum resolved measurements between a single photon and a thermal photon, the latter coming from one of two spatially displaced incoherent sources. We have also showed that such quantum interference phenomena can be exploited to achieve a constant superresolution precision in the estimation of any value of the displacement between the two thermal sources, including the subdiffraction regime where the Rayleigh criteria fails. Interestingly, this is possible without the need of accurately experimentally reproduce the probability distribution at the interferometer output, but by performing simple momentum-resolved sampling measurements from such distribution. Remarkably, it is possible to approach, apart from a factor of 2, the ultimate quantum precision already with a number  of experimental samples of  the order of 2000. Our scheme has an advantage related to its simplicity, allowing superresolution even without the need to decompose the incoming photons modes, which as a consequence implies it is not effected by crosstalk, or requiring their separation to be small, but instead maintaining its efficiency at all scales and doing so by employing only two cameras. Moreover we also showed how it is possible to replace the cameras with two bucket detectors and still achieve superresolution in the regime of small separations.
The advantages in experimental feasibility of this technique together with the possibility to monitor any change in the distance between two incoherent sources can pave the way to important applications in microscopy, astronomy and remote sensing. 

\section*{ACKNOWLEDGMENTS}

We thank Paolo Facchi for the helpful discussions. This project was partially supported by the Air Force Office
of Scientific Research under award number FA8655-23-1-
7046.

\bibliography{bibliography}

\onecolumngrid

\appendix
\section{Probability in Eq.~\eqref{eq:Prob}}
\label{app:Prob}

Here we derive the expression for the probability $P(\Delta K,X)$ in Eq.~\ref{eq:Prob} in the main text, of measuring the photons with a difference in their transverse momenta, $\Delta K=k-k'$, when they hit the same camera ($X=B$, bunching event) or different cameras ($X=A$, antibunching event).
To start we first need to evaluate the second-order correlation function
\begin{equation}
G^{(2)}_{C,C'}(y;y')=\mathrm{Tr}[\hat{\rho}\ \hat{E}_{C}^{-}(y)\hat{E}_{C'}^{-}(y')\hat{E}_{C}^{+}(y)\hat{E}_{C'}^{+}(y')],
\label{eq:appG2Init}
\end{equation}
describing when the two photons, prepared in the initial state expressed by the total density operator expressed by the tensor product $\hat{\rho}=\frac{1}{2}\ket{\psi_{0}}\bra{\psi_{0}}\otimes\left(\ket{\psi_{1}}\bra{\psi_{1}}+\ket{\psi_{2}}\bra{\psi_{2}}\right)$, are detected in the positions $y,y'$ at the cameras $C,C'=C_{1},C_{2}$, and
\begin{equation}
\hat{E}^{+}_{C}(y)=\frac{1}{\sqrt{2\pi}}\sum_{S=0,1}\int \dd x_{s}\ g(x_{s},S;y,C)\hat{a}_{s}(x_{s})
\end{equation}
is the positive frequency part of the electric field operator at the position $y$ of the camera $C$ and $E^{-}_{C}(y)=(E^{+}_{C}(y))^{\dagger}$ is the corresponding negative frequency part,
\begin{equation}
g(x,S;y,C)=\frac{e^{\mathrm{i}\phi(S,C)}}{\sqrt{2}}e^{iK_{0}d}\mathrm{e}^{-i\frac{K_{0}}{\mathrm{d}}xy}
\end{equation}
is the Fraunhofer transfer function describing the propagation of the photons created by the creation operator $\hat{a}^{\dagger}_{S},\  S=0,1$ from initial position $x$ through the beam-splitter and arriving at the camera $C_{1},C_{2}$ in the position $y$ in the far field, and $\phi(S,C)$ is a phase acquired through the beam-splitter.
Given the linearity of the trace we can write Eq.~\eqref{eq:appG2Init} as
\begin{equation}
G^{(2)}_{C,C'}(y;y')=\frac{1}{2}\sum_{i=1,2}\left\{\mathrm{Tr}\left[\ketbra{\psi_{0}}\bigotimes\ketbra{\psi_{i}}\hat{E}_{C}^{-}(y)\hat{E}_{C'}^{-}(y')\hat{E}_{C}^{+}(y)\hat{E}_{C'}^{+}(y')\right]\right\}.
\end{equation}
We can thus define
\begin{equation}
    \hat{E}_{SC}^{+}(y)=\frac{1}{\sqrt{2\pi}}\int \dd x_{S} \  g(x_{S},S;y,C)\hat{a}_{S}(x_{S})\implies \hat{E}^{+}_{C}(y)=\sum_{S}\hat{E}^{+}_{SC}(y),
\end{equation}
so that we can write
\begin{equation}
    G^{(2)}_{C,C'}(y;y')=\frac{1}{2}\sum_{i=1,2}\left\{\sum_{S_{1},S_{2},S_{3},S_{4}}\mathrm{Tr}\left[\ketbra{\psi_{0}}\bigotimes\ketbra{\psi_{i}}\hat{E}_{S_{1}C}^{-}(y)\hat{E}_{S_{2}C'}^{-}(y')\hat{E}_{S_{3}C}^{+}(y)\hat{E}_{S_{4}C'}^{+}(y')\right]\right\}.
\end{equation}
It's possible to see that for $S_{1}=S_{2}$ or $S_{3}=S_{4}$, the contributions above are zero.
This allows us to rewrite 
\begin{equation}
   G^{(2)}_{C,C'}(y;y')=\frac{1}{2}\sum_{i=1,2}  \left|\sum_{S_{1}=0,1}\bra{\mathrm{vac}}E^{+}_{S_{1},C}(y)E^{+}_{\sigma(S_{1})C'}(y')\ket{\psi_{0}}\ket{\psi_{i}}\right|^{2}, \quad \sigma(0)=1,\,\sigma(1)=0,
\end{equation}
and to proceed we evaluate the scalar products:
\begin{equation}
\begin{aligned}
\bra{\mathrm{vac}}E^{+}_{S_{1},C}(y)E^{+}_{\sigma(S_{1})C'}(y')\ket{\psi_{0}}\ket{\psi_{i}}=&\frac{1}{2\pi}\int \dd x \dd z \dd x' \dd z' \ g(x, S_{1}; y, C)g(z, \sigma(S_{1}); y', C')\\
&\times\psi_{0}(x')\psi_{i}(z')\bra{\mathrm{vac}}\hat{a}_{S_{1}}(x)\hat{a}_{\sigma(S_{1})}(z)\hat{a}^{\dagger}_{0}(x')\hat{a}^{\dagger}_{1}(z')\ket{\mathrm{vac}}\\
=&\frac{1}{2\pi}\int \dd x \dd z \dd x' \dd z' \ g(x, S_{1}; y, C)g(z, \sigma(S_{1}); y', C')\\
&\times\psi_{0}(x')\psi_{i}(z')\left(\delta_{0S_{1}}\delta(x-x')\delta(z-z')+\delta_{1S_{1}}\delta(x-z')\delta(z-x')\right)\\
=&\frac{1}{2\pi}\left(\delta_{0S_{1}}\int \dd x \dd z \ g(x, 0; y, C)g(z, 1; y', C')\psi_{0}(x)\psi_{i}(z)\right.\\
&\left.+\delta_{1S_{1}}\int \dd x \dd z \ g(x, 1; y, C)g(z, 0; y', C')\psi_{0}(z)\psi_{i}(x)\right).
\end{aligned}
\end{equation}
It is now possible to rewrite these integrals via the Fourier transform of the spatial wavepackets $\psi_{i}(x)$. If we consider for example the reference source wavepacket:
\begin{equation}
\int \dd x g(x,0;y,C)\psi_{0}(x)=\frac{1}{\sqrt{2\pi}}\frac{\mathrm{e}^{\mathrm{i}\Phi(0,C)}}{\sqrt{2}}\mathrm{e}^{\mathrm{i}K_{0}d}\int\dd x \ \mathrm{e}^{-\mathrm{i}\frac{K_{0}}{d}xy}\psi_{0}(x-x_{0})=\frac{\mathrm{e}^{\mathrm{i}\Phi(0,C)}}{\sqrt{2}}\mathrm{e}^{\mathrm{i}K_{0}d}\mathrm{e}^{-\mathrm{i}\frac{K_{0}}{d}yx_{0}}\phi\left(y\frac{K_{0}}{d}\right),
\end{equation}
where $\phi(\cdot)$ is the Fourier transform in the momentum space of the spatial wavepacket.
Through the right substitution we get the explicit expressions for the correlation functions:
\begin{equation}
G^{(2)}_{C,C'}(y;y')=\frac{1}{2}|\phi(k)|^{2}|\phi(k')|^{2}\left\{1+\frac{1}{2}\mathrm{Re}\left[\mathrm{e}^{\mathrm{i}(\Phi(0,C)+\Phi(1,C')-\Phi(1,C)-\Phi(0,C'))}\left(\mathrm{e}^{-\mathrm{i}(k-k')\Delta x_{1}}+\mathrm{e}^{-\mathrm{i}(k-k')\Delta x_{2}}\right)\right]\right\}
\end{equation}
where we defined the parameters $\Delta x_{1}=x_{0}-x_{1}$, $\Delta x_{2}=x_{0}-x_{2}$ and $k= y \frac{K_0}{d}$, $k'=y' \frac{K_0}{d}$ are the detected transverse momenta of the photons.
Now we can sum these expressions to find the probabilities relative to a bunching or an antibunching event.
Since the beam-splitter action is unitary, the equalities $\Phi(0,C)+\Phi(1,C')-\Phi(1,C)-\Phi(0,C')=+\pi$ for bunching and $-\pi$ for antibunching hold respectively.
Using these equalities, we can find the probability for an antibunching event
\begin{equation}
    \begin{aligned}
        P(k,k',A)=G^{(2)}_{12}(y;y')=
        \frac{1}{2}|\phi(k)|^{2}|\phi(k')|^{2}\left[1-\frac{1}{2}\left(\cos[(k-k')\Delta x_{1}]+\cos[(k-k')\Delta x_{2}]\right)\right],
    \end{aligned}
\end{equation}
and the probability for a bunching event
\begin{equation}
    \begin{aligned}
       P(k,k',B)=\frac{G^{(2)}_{11}(y;y')+G^{(2)}_{22}(y;y')}{2}=
        \frac{1}{2}|\phi(k)|^{2}|\phi(k')|^{2}\left[1+\frac{1}{2}\left(\cos[(k-k')\Delta x_{1}]+\cos[(k-k')\Delta x_{2}]\right)\right].
    \end{aligned}
\end{equation}
We now introduce the variables $K=(k+k')/2$ and $\Delta K=k-k'$, and we integrate over $K$ to obtain
\begin{equation}
P(\Delta K, X)=\frac{1}{2}C(\Delta K)\left[1+\alpha(X)\frac{1}{2}\left(\cos\left(\Delta K \Delta x_{1}\right)+\cos\left(\Delta K \Delta x_{2}\right)\right)\right],
\label{eq:appP1}
\end{equation}
where $\alpha(X)=+1$ for a bunching event and $\alpha(X)=-1$ for antibunching respectively and we defined the envelope function $C(\Delta K)=\int dK \left|\phi(K+\frac{\Delta K}{2})\right|^{2}\left|\phi(K-\frac{\Delta K}{2})\right|^{2}$.
After the change of parameters $\Delta x=x_{1}-x_{2}$ and $x_s=(x_{1}+x_{2})/2$, it is possible to write Eq.~\eqref{eq:appP1} as shown in Eq.~\eqref{eq:Prob} in the main text, i.e.
\begin{equation}
P(\Delta K, X)=\frac{1}{2}C(\Delta K)\left\{1+\alpha(X) \cos\left(\Delta K\Delta x/2\right)\cos\left[\Delta K\left(x_{0}-x_s\right)\right]\right\}
\end{equation}

\section{Fisher Information Matrix}
\label{app:FI}
In this section we will evaluate the Cram\'er-Rao bound shown in Eq.~\eqref{eq:CRB} with the Fisher information shown in Eq.~\eqref{eq:Fisher}.
To do so, we will initially assume that both the parameters $\Delta x$ and $x_s$ associated with the position of the two thermal sources are unknown, and show that the estimation of $\Delta x$ is statistically independent from the estimation of $x_s$ for an aligned setup with $x_0-x_s=0$, and that in this case evaluate the Fisher information $F(\Delta x)$.

The elements of a generic Fisher information matrix $F$ associated with the estimation of any number $p$ of parameters $\boldsymbol{\varphi}$ from a probability distribution $P(\boldsymbol{x};\boldsymbol{\varphi})$ can be calculated as~\cite{Cramer1999}
\begin{equation}
F_{ij}=\int \dd\boldsymbol{x}\ \frac{1}{P(\boldsymbol{x};\boldsymbol{\varphi})}\left(\frac{\partial}{\partial\varphi_i}P(\boldsymbol{x};\boldsymbol{\varphi})\right)\left(\frac{\partial}{\partial\varphi_j}P(\boldsymbol{x};\boldsymbol{\varphi})\right),\quad i,j\in\{1,\dots,p\}.
\label{eq:appFiDef}
\end{equation}
In our estimation problem there are two parameters, $\Delta x$ and $x_s$, so the Fisher information matrix is a 2x2 matrix, of which we can derive the elements applying the expression of the probability in Eq.~\eqref{eq:Prob} to Eq.~\eqref{eq:appFiDef}.
The diagonal term relative to the estimation of $\Delta x$ reads
\begin{equation}
F_{11}=\sum_{X=A,B}\int \dd\Delta K \ P(\Delta K,X)\left\{\frac{\dd}{\dd\Delta x}\mathrm{log}\left[P\left(\Delta K, X\right)\right]\right\}^{2}=\int\dd\Delta K\ f_{11}(\Delta K;\Delta x, x_s).
\end{equation}
As we evaluate the derivative to be
\begin{equation}
\frac{\dd}{\dd\Delta x}P(\Delta K,X)=\frac{1}{2}C(\Delta K)\alpha(X)\left(-\frac{\Delta K}{2}\right)\sin\left(\Delta K\Delta x/2\right)\cos\left[\Delta K\left(x_{0}-x_s\right)\right],
\end{equation}
we can ultimately write the integrand $f_{11}(\Delta K;\Delta x, x_s)$ as
\begin{equation}
f_{11}(\Delta K;\Delta x, x_s)= C(\Delta K)\frac{(\Delta K)^{2}}{4}\frac{\sin^{2}(\Delta K\Delta x/2)\cos^{2}[\Delta K(x_{0}-x_s)]}{1-\cos^{2}(\Delta K\Delta x/2)\cos^{2}[\Delta K(x_{0}-x_s)]}.
\end{equation}
Similarly, for the off-diagonal terms we obtain
\begin{equation}
F_{12}=F_{21}=\int \dd\Delta K \ f_{12}(\Delta K;\Delta x, x_s),
\end{equation}
evaluating the derivative in respect to the centroid
\begin{equation}
\frac{\dd}{\dd  x_{s}}P(\Delta K,X)=\frac{1}{2}C(\Delta K)\alpha(X)\Delta K \cos(\Delta K\Delta x/2)\sin\left[\Delta K\left(x_{0}-x_{s}\right)\right]
\end{equation}
the integrand is:
\begin{equation}
f_{12}(\Delta K;\Delta x, x_s)=- C(\Delta K)\frac{(\Delta K)^{2}}{8}\frac{\sin(\Delta K\Delta x)\sin[2\Delta K(x_{0}-x_s)]}{1-\cos^{2}(\Delta K\Delta x/2)\cos^{2}[\Delta K(x_{0}-x_s)]}
\end{equation}
And last for the element relative to the centroid $x_s$
\begin{equation}
F_{22}=\int \dd\Delta K \ f_{22}(\Delta K;\Delta x, x_s)
\end{equation}
with:
\begin{equation}
f_{22}(\Delta K;\Delta x, x_s)= C(\Delta K)(\Delta K)^{2}\frac{\cos^{2}(\Delta K\Delta x/2)\sin^{2}[\Delta K(x_{0}-x_s)]}{1-\cos^{2}(\Delta K\Delta x/2)\cos^{2}[\Delta K(x_{0}-x_s)]}.
\end{equation}

If we now assume that the position $x_0$ of the reference photon is aligned with the centroid of the two thermal sources, i.e. for $x_0=x_s$, the only non-vanishing integrand is $f_{11}(\Delta K;\Delta x, x_s)$, so that the Fisher information matrix reduces to a diagonal matrix
\begin{equation}
F=\begin{pmatrix}
\frac{1}{4}\int\dd\Delta K\ C(\Delta K)\Delta K^2 & 0\\ 0 & 0
\end{pmatrix}=
\begin{pmatrix}
\frac{\sigma_k^2}{2} & 0\\ 0 & 0
\end{pmatrix},
\end{equation}
where $\sigma_k^2$ is the variance of the transverse-momentum distribution of the photons.
This means that the estimation of $\Delta x$ is statistically independent of $x_s$, and the Cram\'er-Rao bound for the estimation of $\Delta x$ coincides with the Fisher information in Eq.~\eqref{eq:Fisher}.

\section{Non-resolving detectors}
\label{app:Nonres}

The already discussed experimental set-up includes two resolving cameras at the outputs of the beam splitter, now we consider to replace them with two bucket detectors which cannot measure the transverse momenta of the photons $k$,$k'$.
The only information we can get is whether the photons are detected by the same ($X=B$) or different ($X=A$) detectors, happening with probabilities:
\begin{equation}
P(X)=\int \dd \Delta k \ P(\Delta K, X)= \frac{1}{2}\left\{1+\alpha(X)\int \dd \Delta K \ C(\Delta K)\cos\left(\Delta K\frac{\Delta x}{2}\right)\right\},
\end{equation}
obtained by the probability in Eq. (\ref{eq:Prob}) aligning the reference source ($x_{0}=2x_{s}$) and integrating.\\
So it is now possible to evaluate the Fisher Information associated to the estimation of $\Delta x$ when we do not employ a measurement resolving the transverse momenta of the photons. So we sum over the two possible outcomes:
\begin{equation}
\begin{aligned}
F^{(nr)}(\Delta x)=\sum_{A,B}P(X)\left(\frac{\dd}{\dd \Delta x}\mathrm{log}P(X)\right)^{2}&=\sum_{A,B}\frac{1}{2}\frac{\left(\int \dd \Delta K\ C(\Delta K)\frac{\Delta K}{2}\sin(\Delta K\frac{\Delta x}{2})\right)^2}{\left\{1+\alpha(X)\int \dd \Delta K \ C(\Delta K)\cos\left(\Delta K\frac{\Delta x}{2}\right)\right\}}\\
&=\frac{1}{4}\frac{\left(\int \dd \Delta K\ C(\Delta K)\Delta K\sin(\Delta K\frac{\Delta x}{2})\right)^2}{1-\left(\int\dd \Delta K\ C(\Delta K)\cos\left(\Delta K\frac{\Delta x}{2}\right)\right)^{2}}.
\end{aligned}
\end{equation}
In the regime of small separations $\sigma_{x}\Delta x\ll 1$, in which the wavepackets are mostly overlapping, we can expand neglecting the higher orders in $\Delta x$, obtaining:
\begin{equation}
F(\Delta x)\simeq \frac{1}{4}\frac{\left(\frac{\Delta x}{2}\right)^{2}}{\left(\frac{\Delta x}{2}\right)^{2}}\frac{\left(\int\dd \Delta K\ C(\Delta K) \Delta K^{2}\right)^{2}}{\int\dd \Delta K\ C(\Delta K) \Delta K^{2}}=\frac{\sigma^{2}_{k}}{2},
\end{equation}
which is the same as the FI in Eq. \ref{eq:Fisher} in the main text. So in the case of small separations, $\sigma_{x}\Delta x\ll 1$, it is possible to be in the superresolution regime even using two bucket detectors in the place of resolving cameras.

\end{document}